\documentclass[aps,prstab,twocolumn,groupedaddress,amsmath,nobibnotes,nofootinbib]{revtex4-1}

\usepackage{graphicx}
\usepackage{subfigure}
\usepackage{tabularx}
\usepackage{threeparttablex}
\usepackage{natbib}
\usepackage[outdir=./]{epstopdf}

\newcommand\Tstrut{\rule{0pt}{2.6ex}}         
\newcommand\Bstrut{\rule[-0.9ex]{0pt}{0pt}}   

\begin{document}


\title{Quasi-integrable optics for a small electron recirculator}


\author{K. Ruisard,\textsuperscript{1}\email{ruisardkj@ornl.gov} H. B. Komkov,\textsuperscript{2} B. Beaudoin,\textsuperscript{2} I. Haber,\textsuperscript{2} D. Matthew,\textsuperscript{2} and T. Koeth\textsuperscript{2}}
\affiliation{\textsuperscript{1}Oak Ridge National Laboratory, Oak Ridge, Tennessee 37830, USA}
\affiliation{\textsuperscript{2}Institute for Research in Electronics and Applied Physics, College Park, Maryland 20742, USA}

\date{\today}

\begin{abstract}
This paper describes the design and simulation of a proof-of-concept quasi-integrable octupole lattice at the University of Maryland Electron Ring (UMER). 
This experiment tests the feasibility of nonlinear integrable optics, a novel technique that is expected to mitigate resonant beam loss and enable low-loss high-intensity beam transport in rings. 
Integrable lattices with large amplitude-dependent tune spreads, created by nonlinear focusing elements, are proposed to damp beam response to resonant driving perturbations while maintaining large dynamic aperture [Danilov and  Nagaitsev, \textit{PRSTAB}, 2010]. 
At UMER, a lattice with a single octupole insert is designed to test the predictions of this theory.
The planned experiment employs a low-current high-emittance beam with low space charge tune shift ($\sim0.005$) to probe the dynamics of a lattice with large externally-induced tune spread. 
Design studies show that a lattice composed of a 25-cm octupole insert and existing UMER optics can induce a tune spread of $\sim 0.13$. 
Stable transport is observed in PIC simulation for many turns at space charge tune spread $0.008$. A maximum spread of $\Delta \nu = 0.11$ (RMS $0.015$) is observed for modest octupole strength (peak $50\ T/m^3$).
A simplified model of the system explores beam sensitivity to steering and focusing errors. Results suggest that control of orbit distortion to $<0.2$ mm within the insert region is essential. However, we see only weak dependence on deviations of lattice phase advance ($\leq0.1$ rad.) from the invariant-conserving condition. 
\end{abstract}


\maketitle


\section{Introduction}
One of several aspects limiting transportable beam current in accelerators is uncontrolled beam loss. 
Even low fractional losses in an intense beam presents a hazard to personnel safety and the integrity of accelerator components as well as compromising beam quality. 
Known loss mechanisms include both incoherent and coherent resonances that drive particles to large amplitudes or lead to beam instability. 
This is especially of concern in high-intensity machines operating with large space charge tune spreads, which may encompass many periodic orbits of low order. 

Moderate tune spreads have been known to stabilize beams against resonant interactions. In particular, nonlinear terms in the transverse focusing force can create a beneficial amplitude-dependent tune spread. 
The best known example is octupole-induced Landau damping, in which an octupole-induced tune shift in the particle distribution damps transverse collective instability \cite{Chao}.
In the presence of strong nonlinearity, regular driving terms cannot resonantly couple energy into the beam, as collective motions decohere and driven orbits shift away from the resonant condition \cite{Webb2013}. 
The difficulty with imposing large tune spreads is the associated loss of dynamic aperture due to the emergence of chaotic orbits near resonance overlap. 
A solution has been proposed by Danilov and Nagaitsev in the theory of nonlinear integrable optics (NLIO) \cite{Danilov2010}. In this theory, integrable (non-chaotic) orbits are maintained for arbitrarily strong nonlinear focusing with large amplitude-dependent tune shifts. Reference \cite{Danilov2010} identifies a family of nonlinear magnetic potentials in which transverse particle orbits conserve coupled, quadratic invariants of motion that are distinct from the Courant-Snyder invariants. 

NLIO is a potential pathway for future high-intensity machine design, in which the traditional linear (quadrupole) optics are combined with highly-nonlinear focusing elements. 
These nonlinear-focusing rings are predicted to be much more robust against single-particle resonances and coherent instability. 
Many numerical studies have been performed to gauge the feasibility of the NLIO concept, including the the effect of lattice errors, dispersion and chromaticity on integrability \cite{Webb2014,Nagaitsev2012}.
More recently, discussion has extended to understanding the contribution of space charge to invariants and beam stability \cite{Webb2014,Hall2016}.
A design study for a NLIO rapid-cycling synchrotron has shown robust performance with space charge tune shifts $\Delta \nu=0.05$ \cite{Eldred2018}.
However, at this point in time the concept is awaiting demonstration. 
Progress towards this goal is measured in the commissioning of the Integrable Optics Test Accelerator (IOTA), described in reference \cite{Antipov2017}, as well as the effort described here.

An experimental program at the University of Maryland Electron Ring (UMER) has been developed as a testbed for the NLIO concept. 
The experiment is based on the theory of quasi-integrable optics (QIO), in which only one invariant of transverse motion is preserved. Compared to NLIO, the maximum stable tune spread induced on the particle distribution is expected to be smaller, but otherwise the lattice is similar in design and expected behavior. 
This program benefits from the ability of UMER to operate at a wide range of space-charge densities, which may provide insight into operation with space-charge beams at and beyond the previously studied concentrations.

The design of a strongly nonlinear lattice is in itself a challenge, as there is no operational experience to draw on and the standard principles for accelerator design must be extended or modified.  
In the case of UMER, an additional constraint is found in seeking to implement an integrable optics design as an upgrade to an existing quadrupole lattice. 
We believe insights developed in retrofitting the UMER FODO lattice for QIO operation, including a design for lightweight printed-circuit octupole insert and studies of predicted error sensitivity, may serve as a guide for future rings with integrable or  quasi-integrable optics. 

Section \ref{sec:umer} reviews the basic parameters of the UMER lattice. 
Section \ref{sec:design} is an overview of the design of a quasi-integrable octupole lattice for UMER. This section describes both development of the octupole insert as well as optimization of the linear-focusing optics for QI transport and the available beam parameters for experiments. 
In Section \ref{sec:dynamics}, particle-in-cell (PIC) simulations are used to explore particle dynamics in the proposed lattice. Conclusions are drawn regarding optimal operating point and required tolerances.
Finally, Section \ref{sec:discuss} provides discussion of the lessons learned and outlook for the UMER QIO program.

\section{UMER Program} \label{sec:umer}

UMER is a scaled, 10 keV ($\beta = 0.195$) electron ring experiment for the study of high-intensity beam dynamics relevant to higher-energy ion rings. A range of space charge densities are selected by aperturing the beam near the source, in the range $ \nu/ \nu_0 = 0.95 \text{ to } 0.14$ for nominal tune 6.7 (incoherent shifts $\Delta \nu = 0.3 \text{ to } 5.7$) \cite{Kishek2014,Bernal2017}. As such, typical operation in UMER exceeds the space charge intensity of existing, production-level rings such as $\nu/\nu_0=0.98, \Delta \nu = 0.15$ in the SNS accumulator ring \cite{Wei2000}.
 
The built-in flexibility of UMER is easily leveraged for QIO experiments. The linear optics are highly configurable thanks to independently-powered quadrupoles. The low beam rigidity allows for the use of low-cost, air-core printed circuit board (PCB) magnets; we apply the same approach to create our nonlinear focusing elements.
The lattice also includes induction cells for longitudinal focusing, which supports multiple confinement conditions such as RF sine (recently added) and RF barrier bucket in addition to unconfined/coasting bunches \cite{Hamilton2018,Beaudoin2016}.

In retro-fitting any experiment for a new configuration, there are challenges. For UMER these included the high density of ring optics, large orbit distortion (permissible in typical operation) and the large space-charge tune depression which exceeds the predicted octupole-induced spread. 
Each of these points is expanded on where appropriate below.

\section{Design of a Quasi-integrable Octupole Lattice}  \label{sec:design}

The defining attribute of a NLIO/QIO lattice are nonlinear insertion elements embedded within a linear-focusing lattice. Both the shape of the nonlinear potential and the transfer function of the linear-focusing lattice must be tailored to ensure integrability. 
The recipe laid out in Reference \cite{Danilov2010} is re-iterated here for convenience. In this approach, constant Laplacian potentials are identified that permit one or two invariants of motion. 
The nonlinear potential is constant in normalized phase space coordinates \footnote{$x_N \equiv \frac{x}{\sqrt{\beta_x(s)}}$, $p_{x,N} \equiv p_x\sqrt{\beta_x(s)}+\frac{\alpha_x x}{\sqrt{\beta_x(s)}}$} and therefore has lab-frame longitudinal dependence that scales with the betatron function $\beta(s)$, where exact scaling depends on pole order in the multipole expansion of the nonlinear potential.
For proper cancellation of terms, the beam must be round through the nonlinear insert, $\beta_x(s)=\beta_y(s)$. 
In implemention, linear-focusing ring sections are tailored to form a round waist at the insert location.
Finally, for quasi-continuous motion through the nonlinear potential the phase advance in the linear-focusing optics between inserts must be integer-$\pi$. 
This of course means that any additional contribution to particle trajectories, such as space charge depression or off-momentum effects, causes orbits to shift away from the known invariant \cite{Webb2014}.

\subsection{Nonlinear Insert} \label{sec:octu}

\begin{figure}[tb]
\centering
\subfigure[ Left: Half of printed circuit octupole magnet. Right: top half of 25 cm composite octupole channel containing seven 4.65 cm PCBs. ]{
\hspace{0.01\textwidth}
\includegraphics[width=0.35\textwidth]{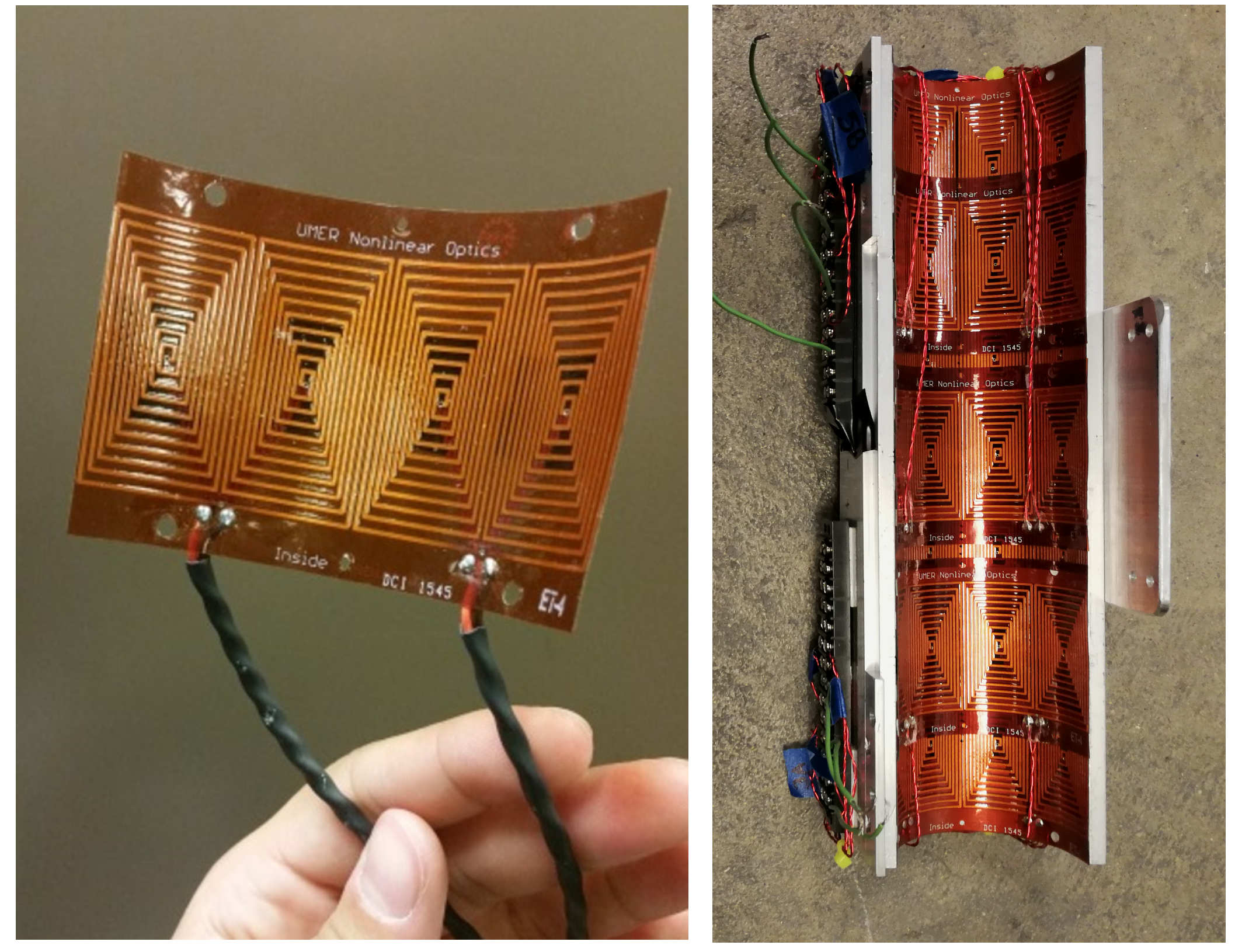}   
\label{fig:octu-pcb}
}
\subfigure[ Longitudinal profile of composite octupole channel with over-lapping, evenly-spaced PCBs. Profile is determined from Biot-Savart integration of the designed circuit.]{
\includegraphics[width=0.45\textwidth]{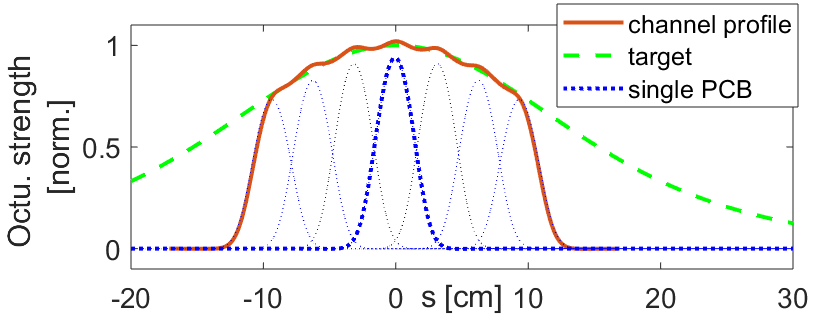}   
\label{fig:octu-channel}
}
\caption{Printed circuit octupole magnets and long octupole channel designed for QIO experiments. }
\label{fig:octu}
\end{figure}  

An early choice for the UMER lattice was to test quasi-integrable, rather than integrable, optics. This is motivated by the predicted relaxed tolerances of the QIO system (compared to NLIO) and the significantly reduced complexity in designing a single-multipole magnetic potential. In particular, the UMER PCB magnets can easily be configured for any $\cos{n\theta}$ current distribution \cite{Zhang2000}. \footnote{This is in contrast to the IOTA insertion device, which uses shaped iron pole-pieces for both a QI octupole insert and a fully integrable potential \cite{Antipov2017,OShea2015}.}

In the case of octupole fields, the single invariant of transverse motion takes the form (in normalized canonical coordinates) \cite{Danilov2010}
\begin{align}
H_N = &\frac{1}{2} \left( p_{x,N}^2 + p_{y,N}^2 +x_N^2 + y_N^2 \right) \nonumber \\
&+ \frac{\kappa}{4} \left(x_N^4 + y_N^4 - 6y_N^2x_N^2\right).
\label{eq:H-norm}
\end{align}
\noindent Following from the the $x^4$ dependence of the fields, the octupole strength parameter $\kappa$ is s-independent (and $H_N$ is invariant) if the octupole term has dependence proportional to $\beta^{-3}(s)$.

\begin{figure}[t]
\includegraphics[width=0.45\textwidth]{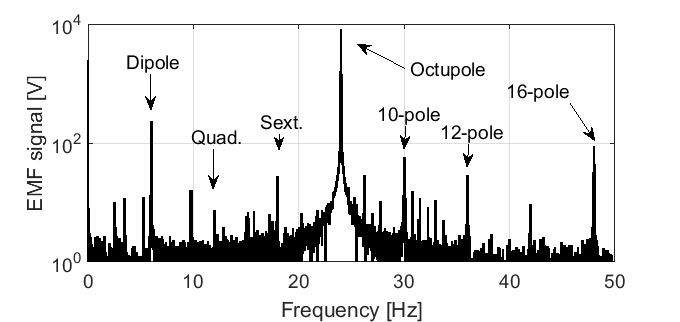}   
\caption{ FFT from rotating coil measurement of single (short) octupole magnet. Large dipole peak is due to presence of ambient Earth magnetic field and does not originate from octupole circuit.}
\label{fig:octu-fft}
\end{figure}

The nominal UMER FODO lattice is very dense (magnet fill factor $43\%$), due to the need to provide confinement for very high intensity beams. With $36$ bends in an $11.52$ meter circumference, the longest straight section available for an insertion device is $25$ cm ($32$ cm minus mechanical clearances). 
To maximize configurability, short 4.65 cm-long octupoles were fabricated as flexible PCBs (Fig. \ref{fig:octu-pcb}) and assembled into a composite ``octupole channel" consisting of seven independently-powered octupole circuits \cite{BaumgartnerNAPAC2016, BaumgartnerIPAC2018}. 
Due to the thin PCB design, these elements overlap within the channel to produce a fairly smooth longitudinal profile $K_3(s)$ as shown in Fig. \ref{fig:octu-channel}. 
There is $2\%$ RMS flutter about the ideal profile $\propto \beta(s)^{-3}$, but numerical studies predict minimal effect on dynamic aperture. 
Individual PCB octupoles have a peak gradient strength $51.6 \pm 1.5 \text{T}/\text{m}^3/\text{A}$ and unwanted multipole orders are suppressed by two orders of magnitude as measured by a rotating coil, shown in Fig. \ref{fig:octu-fft}. 

\subsection{Choice of operating point}

The fractional tune of the lattice will be equal to the total ``tune advance" $\nu_{ch}$ through nonlinear insertions (or equal to $\frac{1}{2}-\nu_{ch}$). 
This fractional tune should be chosen to maximize tune spread; it is observed in simulation that maximum induced tune spread is approximately equal to the fractional tune, that is $\Delta \nu_x = \Delta \nu_y \sim \nu_{ch}$. 
Therefore, largest possible tune advance is desired, which corresponds with smallest possible beam waist in the insert region (therefore, the design is similar to that of mini-beta insertions for colliders). Here we use notation $\beta_*$ to indicate the betatron function at the waist.
To avoid scraping of a $100\ \mu$m beam in the quadrupoles flanking the insert region, $\beta_*>0.23$ m is required (using a $2\times\text{edge}$ safety factor from the 5 cm pipe diameter). For the UMER design, $\beta_*=0.3$ m is identified as a feasible target for largest possible phase advance (which correlates with maximizing the octupole damping term) that can be supported by existing optics. 
For a single 25-cm octupole insert, the ring operating point should be $\nu_x=\nu_y=0.13$, and the maximum tune spread (across all stable orbits) is $\sigma_{\nu} = 0.13$.

The octupole strength parameter $\kappa$ should be chosen to maximize amplitude-dependent tune spread but also ensure sufficient dynamic aperture to avoid particle losses. 
In the octupole lattice, there is a sharp edge to the dynamic aperture and the induced tune shift is largest for particles near the aperture limit. 
This limit shrinks with increased octupole strength and is bounded by unstable fixed points in the quasi-integrable potential at $\left(x_N,y_N\right)=\left(\pm 1/\sqrt{2\kappa}, \pm 1/\sqrt{2\kappa}\right)$, first described in reference \cite{Webb2013}. 
Therefore, an estimate for maximum stable aperture is $r_{max} = \sqrt{\beta_*/\kappa}$. 

For the 25-cm insert over a $\beta_*=0.3$ m waist, a peak octupole gradient of $50\ T/m^3$ places the fixed point at $3.2\times\text{RMS}$ radius for a $100$ mm-mrad $4\times$RMS emittance beam. 
This requires only 0.97 A in the central octupole circuit, well within the safety limit of UMER PCB octupoles.

\subsection{Linear-focusing Optics}

\begin{figure}[tb]
\centering
\includegraphics[width=0.5\textwidth]{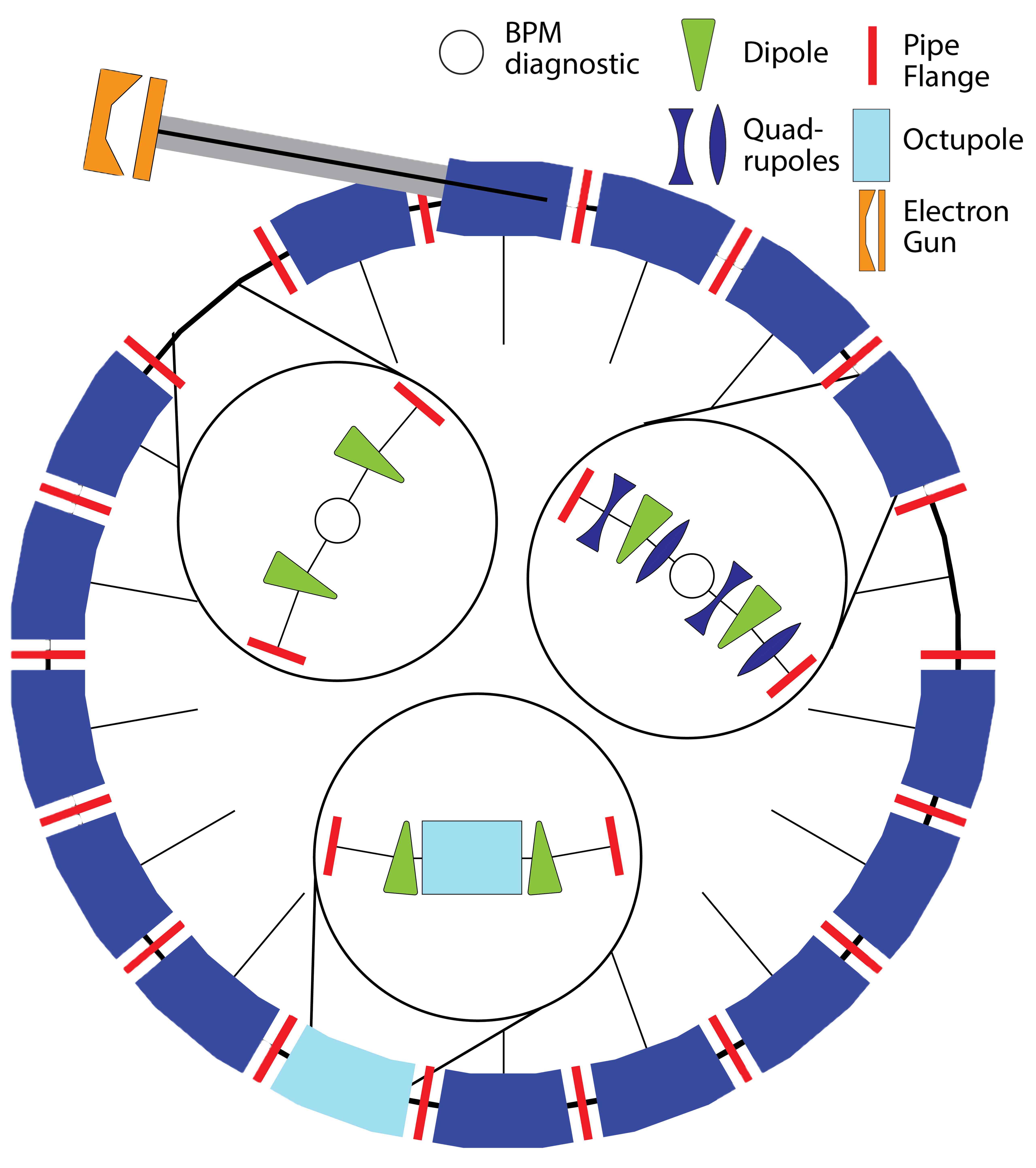}
\caption{Diagram of UMER QIO lattice, with insets showing detail for standard $20^\circ$ section (Q-D-Q-Q-D-Q) and octupole section (0-D-channel-D-0). Spokes indicate diagnostic locations and red bands indicate vacuum flanges. }
\label{fig:exp-layout-a}
\end{figure}

\begin{table}[htb]
\centering
\caption{Nominal parameters for the UMER QIO lattice. }
\label{tab:umer-params}
\begin{ruledtabular}
\begin{tabular}{l @{\extracolsep{\fill}} cc}
Parameter & Value &   \Tstrut\Bstrut\\
\hline  
Beam energy & 10 keV&    \Tstrut\\
Beam current & $0.04 \text{--} 0.6$ mA &\\
Incoh. $\Delta \nu$ & $0.005 \text{--} 0.94$ &\\
Circumference & 11.52 m &\\
Pipe diameter & 5.08 cm &\\
Bunch length &  $20 \text{--} 100$ ns  &\\
Super-period & 3.84 m &\\
Quadrupoles per period & 20 &\\
Bare tune ($\nu_x = \nu_y$) & 3.13 ($3.1 \text{--}  3.4$ range)&\\
Natural chromaticity & $C_x = -4.3$ &\\
		& $C_y = -3.2$&\\
\end{tabular}
\end{ruledtabular}
\end{table}

\begin{figure}
\centering
\includegraphics[width=0.5\textwidth]{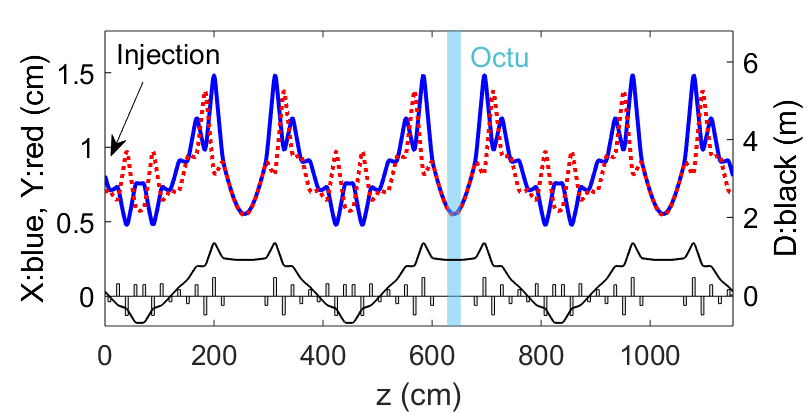}
\caption{\label{fig:lattice-solution} Periodic KV envelope solution for 100 mm-mrad, 60 $\mu$A beam at $\nu_x=3.124$, $\nu_y=3.128$. Injection is at s=0 plane..}
\end{figure} 

The layout of the UMER QIO experiment is shown in Fig. \ref{fig:exp-layout-a}, while Table \ref{tab:umer-params} summarizes the parameters for the lattice as designed. The design includes a single nonlinear insertion, which will be sufficient for proof-of-principle experiments. 
In general, multiple insertions should lead to stronger resonance damping and room is held in this design for additional octupole inserts with minimal modification.
As the ring is built in modular $20^{\circ}$ arcs, each housing two dipole bends and four quadrupoles (Q-D-Q-Q-D-Q), an entire $20^{\circ}$ section is modified to accomodate the insertion. The long insert occupies the straight between dipoles (0-D-OCTU-D-0), as shown via inset in Fig. \ref{fig:exp-layout-a}.

The only fixed parameters in the linear lattice design are the magnet locations. There are two quadrupoles per dipole (72 total) at equal 16 cm spacing. To make room for a long insert it is necessary to physically remove two quadrupole elements, leaving 70 available for meeting the integer-$\pi$ requirement.
A few-period solution is chosen, shown in Fig. \ref{fig:lattice-solution}, in which 60 of the available quadrupoles are used for the linear focusing in periodic arrangement of 10 unique strengths (optimization problem is reduced to 10 degrees of freedom).  

The choice of a few-period design with many quadrupoles per period allows for flexibility of lattice function (ie, shifting operating points in linear lattice without changing optics of insert region) and supports largely adiabatic changes to the beam size (empirically observed to reduce sensitivity to quadrupole strength errors). In this case, super-periodicity of six is theoretically possible (8 quads per cell, 4 degrees of freedom) but this restricts flexibility in tune operating points.\footnote{Periodicity refers to the nominal linear optics design. Due to modified optics at injection, the true super-periodicity of the ring is one.} Periodicity four is more tractable, but periodicity three is chosen to align with diagnostics (at periodicity eighteen). 

The design shown in Fig. \ref{fig:lattice-solution} assumes a low-current, large-emittance beam ($40\mu$A, 100 mm-mrad 4$\times$RMS, unnorm.) for low space charge tune spread during initial experiments. The tune of the lattice shown is $\nu_x = 3.124$, $\nu_y = 3.128$. This is much lower than the typical operating parameters for UMER ($\nu_x = 6.68$, $\nu_y = 6.69$). The weaker focusing allows for long (64 cm) quadrupole-free sections while avoiding unacceptably large excursions of the beam envelope.

There is some flexibility in choosing the orientation of the three-period lattice solution. Six configurations are possible, as each $120^\circ$ super-period is comprised of six $20^\circ$ modular sections (Q-D-Q-Q-D-Q). The orientiation shown in Fig. \ref{fig:exp-layout-a} is chosen based on two primary considerations: low local closed orbit distortions at the insert location (as measured in typical operation) and injection where the horizontal beam size is small. \footnote{Due to a peculiarity in the UMER injection design,  injection cannot be done in one of the long drifts as an off-axis quadrupole is used for simultaneous focusing and bending of the beam into the ring.}

\subsubsection*{Off-momentum orbits} 
Maintaining momentum acceptance is important for QIO ring design. In general, off-momentum orbits will not meet the conditions for quasi-integrability. 
The effects on a NLIO lattice are discussed in reference \cite{Webb2014}, which recommends that chromaticities be equal and dispersion as low as possible. While the UMER lattice includes neither chromaticity-correcting sextupoles nor a dispersion-matching section at injection, low momentum-deviation is generally expected in coasting beam and RF-confined experiments. 
Unconfined beam from the gridded triode electron source is expected to generally have low energy spread, $\delta p / p \sim 0.001$, except in the eroding head and tail of the bunch when confining fields are off \cite{Beaudoin2014,Cui2004}. 
Energy spread is not expected to increase significantly with RF confinement, as low-charge beams are seen in operation to require very low RF voltages \cite{Hamilton2018}. 

With natural chromaticities $C_x = -4.3$, $C_y = -3.3$, the expected coasting beam chromatic tune spreads are $\sim 0.004$. 
If necessary, this may be corrected by the addition of PCB sextupole elements.
For the matched dispersion in Fig. \ref{fig:lattice-solution} we expect $x_D=0.09$ cm in the octupole for a $+10$ eV error in energy. This leads to additional terms in the invariant $H_N$ (Eq. \ref{eq:H-norm}) that may lead to diffusion. The case for unmatched dispersion has yet to be addressed, but the dependence can be probed in experiment as well as in simulation.

\subsection{Electron beam parameters for QIO experiments} \label{sec:beam-params}

\begin{table}
\centering
\caption{UMER beam parameters. The top three rows are proposed for QIO experiments. $r_0$ is radius of aperture used to form beam, emittance $\epsilon$ is unnormalized RMS emittance, $\nu / \nu_0$ values are estimated for FODO lattice with operating point $\nu_x = \nu_y = 6.8$, and tune shifts are estimated for a KV equivalent beam. }
\label{tab:umer-beams}
\begin{ruledtabular}
\begin{threeparttable}
\begin{tabular}{l @{\extracolsep{\fill}}cccccc}
Current [mA] &$r_0$ [mm] & $\epsilon$ [mm-mrad] & $\nu / \nu_0$ & Incoh. $\Delta \nu$ \Tstrut\Bstrut\\ 
\hline 
0.04 (0.01 - 0.1) & -- & $25 \pm 5$  & 1.00 & -0.005\Tstrut \\
0.06 $\ast$ & 0.7 & $2.5$  & 0.95 & -0.3 \\
 0.6 & 0.25&  7.6 & 0.85 & -0.94 \\
 6.0 &0.875&  25.5 & 0.62 & -2.4  \\
 21  &1.5  &  30.0 & 0.31 & -4.5  \\
 80  &2.85 &  86.6 & 0.17 & -5.5  \\
 100 & 3.2 &  97.3 & 0.14  & -5.7  \\
\end{tabular}
\begin{tablenotes}
\footnotesize
\item $\ast$ see references \cite{Bernal2017} and \cite{Bernal2018}
\end{tablenotes}
\end{threeparttable}
\vspace{-3.5mm} 
\end{ruledtabular}
\vspace{-3mm}
\end{table}

As the invariants found in Reference \cite{Danilov2010} are calculated in the single particle limit, initial tests of the QIO system will focus on beams with weak space charge prior to exploring higher intensity regimes. The parameters for all available UMER beams are given in Table \ref{tab:umer-beams}. The ``standard" UMER beams, generated using a single aperture at the source, occupy rows three and below ($\geq 0.6$ mA). Clearly, the incoherent tune shifts $\geq 0.94$ far exceed the predicted octupole-generated tune spread ($\leq 0.13$). The beams described in rows one and two were developed for the QIO experiments. 

The $60\ \mu$A beam (row two) is generated using an aperture -- solenoid -- aperture system to reduce beam density while maintaining low emittance, and is described in references \cite{Bernal2017,Bernal2018}. The nominally $40\ \mu$A beam (First row in Table \ref{tab:umer-beams}) is generated by operating the UMER triode electron gun in voltage amplification mode \cite{KierstenThesis}. In this mode the output current depends strongly on the bias voltage in the triode, in the range $0.01 - 0.1\ mA$. This beam has low incoherent tune shift due to the large emittance; quadrupole scan slice-emittance measurement at $40\ \mu A$ output current yields $\epsilon_x = 300$ mm-mrad, $\epsilon_y = 100$ mm-mrad (unnorm., $4\times$RMS) at the mid-pulse location. The large asymmetry in this measurement is likely due to the time-varying injection kick or large mismatch as profile measurement near the source indicate that $\epsilon_x \approx \epsilon_y$.

\subsection{Simulation tools}
The performance of the QIO lattice depends on the correct execution of the linear-focusing lattice as well as the precision of the octupole insert. 
PIC simulations are performed using the transverse-slice package in the WARP code \cite{warp}. We use a ``simple model" of the QIO system, consisting only of the octupole element and an ideal thin-lens symmetric-focusing kick as a proxy for the linear focusing sections. Additionally, we implement a full ring model of the QIO experiment configuration, using hard-edged models for the quadrupoles. 
In the ring model used here, dipoles are excluded for simplicity and ease of tuning. The UMER dipoles introduce a vertical focusing component due to the fringe fields and sextupole term in the PCB elements. 
Quasi-integrable lattice solutions including the dipole contributions to linear focusing have been found and appear very similar to the solution shown in Fig. \ref{fig:lattice-solution} (see for example reference \cite{RuisardIPAC2018}) but are not reported on here. 
For the octupole insert, we implement both an ideal octupole potential with smoothly-varying longitudinal profile or gridded field model representative of the octupole channel as designed. The gridded model is generated by integrating the Biot-Savart law for the PCB current distribution using in-house code MagLi \cite{magli}. 

40,000 macro-particles are used for modeling the space charge contribution, while the frequency map is sampled using a zero-charge, gridded ``witness distribution." 
The test case beam is initialized as a $60\ \mu$A semi-Gaussian distribution with $100\ \mu$m edge emittance in both planes. In simulation this is measured to have an incoherent tune spread of $\Delta \nu=-0.008$. This is slightly higher current than the nominal case for the low-current beam (Table \ref{tab:umer-beams}, first row).

\section{Dynamics in octupole lattice}  \label{sec:dynamics}

\begin{figure}
\centering
\subfigure[Tune footprint with up to fourth order resonance lines indicated.]{
\includegraphics[width=0.7\columnwidth]{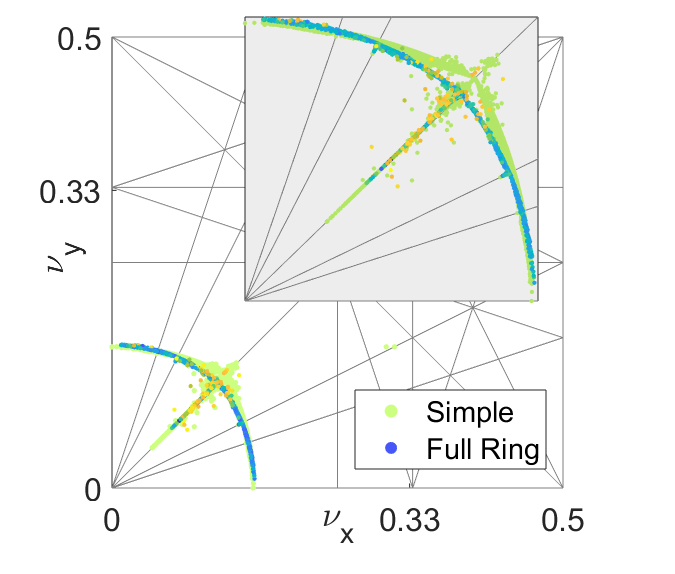}
\label{fig:fma2-tune}
}
\subfigure[Configuration space plot of aperture and resonant structure as a function of particle initial condition. The fixed point at $x=y=\sqrt{\beta_* / 2\kappa}$ is indicated here by the red cross marker.]{
\includegraphics[width=0.7\columnwidth]{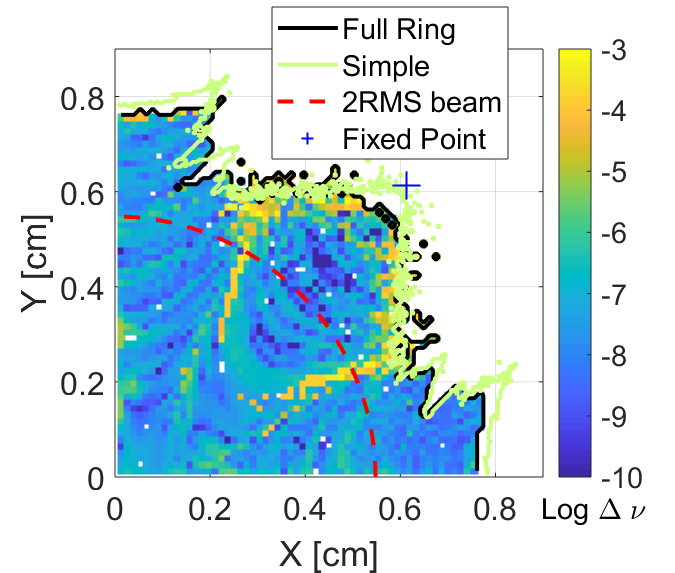}
\label{fig:fma2-xy}
}
\subfigure[Snapshot of beam distribution at beam waist]{
\includegraphics[width=0.7\columnwidth]{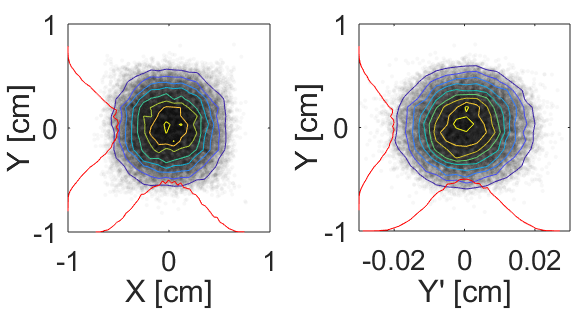}
\label{fig:fma2-dist}
}
\caption{WARP PIC calculation for QIO lattice at peak octupole strength $50\ \text{T}/\text{m}^3$ ($\kappa=3984\ m^{-1}$) and fractional tune 0.13 after 1024 turns.}
\label{fig:fma2}
\end{figure}
\subsection{Transport at design operating point} 

The nominal lattice design, as described above, is for a single-insert lattice with 25-cm octupole insert over  $\beta_*=0.3$ waist, with linear optics optimized to be near the quasi-integrable condition $\nu_x = \nu_y=0.126$.
 Figure \ref{fig:fma2} shows the result of PIC calculations in this case, including the sampled frequency map and final beam distribution.
The dynamic aperture limit can clearly be seen in the frequency map shown in Fig. \ref{fig:fma2-xy}, as well as the characteristic eight-lobed structure imprinted by the octupole potential. 
There is excellent agreement with the equivalent simple model (25-cm channel + thin-lens focusing kick), whose stability boundary is shown as the light green contour in Fig. \ref{fig:fma2-xy}. 
The ``airplane" shape of the tune footprint in Fig. \ref{fig:fma2-tune} is also characteristic, with high-amplitude particles being shifted to the tips of the ``wings." 

The slight tilt in both tune and configuration space is characteristic of a small tune asymmetry in the PIC model, with $\nu_x-\nu_y =-0.001$. 
The simulated tune spreads are $\Delta \nu_x = -0.10$, $\Delta \nu_y = -0.11$ and at the given peak strength $50\ T/m^3$, dynamically unstable orbits appear at $2.27\times$RMS radius. 
The effect of tune depression at the bunch center can be seen in the tune diagram (Fig. \ref{fig:fma2-tune}).
A snapshot of the bunch distribution in Fig. \ref{fig:fma2-dist} shows a slight ``squaring'' in configuration space, which is to be expected in the octupole potential. 

Transport at this operating point is summarized in row 1 of Table \ref{tab:pic-results}. Conservation of the invariant $H_N$ for stable orbits (still confined at 1024 turns) is shown in the last column. In general, orbits exhibit small oscillations $<1\%$ in the zero-charge limit. Including the space charge contribution causes larger oscillations, although orbits still appear to be bounded. Within the first 30 turns, there is a rapid redistribution of charge which is reflected in large changes in single particle amplitudes $H_N$.

\begin{table}[tb]
\centering
\caption{Predicted performance of two tune operating points for PIC simulation of QIO ring with hard-edged elements and peak octupole gradient $50\ T/m^3$ ($\kappa=3984\ m^{-1}$). Tune spreads are calculated as difference from bare tune in linear lattice; the maximum spread is calculated for all stable orbits, while the RMS spread is calculated for all orbits within $2\times$RMS of the bunch center. Conservation of invariant $\text{std} H_N/H_N$ is taken as average over all stable particle orbits within $2\times$RMS beam radius. }
\label{tab:pic-results}
\begin{ruledtabular}
\begin{tabular}{l @{\extracolsep{\fill}}cccc}
Tune 	& $\Delta \nu_x$  	& $\Delta \nu_y$ 	& eff. $r_{max} $ & $\left<\text{std} \left(H_N\right)/H_N\right>$\Tstrut\Bstrut\\ 
		& max/RMS 			& max/RMS 			& 				& 	0 A / $60\ \mu$A					\\
\hline 
0.13	& $ 0.101$/$ 0.015$ 	& $ 0.110$/$ 0.016$	& 0.62 cm 		&0.91\% / 6.54\%					\Tstrut \\
0.26	& $ 0.042$/$ 0.007$	& $ 0.050$/$ 0.005$	& 0.40 cm 		&8.02\% / 24\%					\\
\end{tabular}
\end{ruledtabular}
\end{table}

\subsection{Operation near fourth order resonance}

One thrust of the UMER QIO lattice design was to maximize the induced tune spread, which scales with the fractional tune. 
An earlier version of the design assumed a 64 cm octupole section, in which the central 25 cm insert is flanked on either side by one to two short octupole PCBs, with discontinuities to accomodate the two $10^{\circ}$ bends. 
This was shown in simulation to lead to large loss of stable orbits and this scheme was abandoned. (For more details, refer to references \cite{KierstenThesis} and \cite{RuisardICAP2018}.) 
PIC simulations are still run for this configuration, using the 25 cm insert but with tune $\nu_x=\nu_y=3.263$ optimized for a 64 cm insert over $\beta_*=0.3$ m waist.

\begin{figure}
\centering
\subfigure[Tune footprint with up to fourth order resonance lines indicated.]{
\includegraphics[width=0.7\columnwidth]{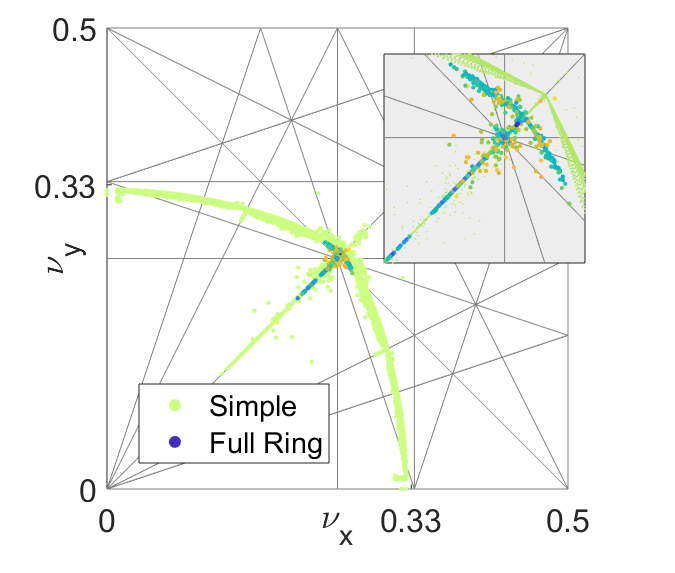}
\label{fig:fma1-tune}
}
\subfigure[Configuration space plot of aperture and resonant structure.]{
\includegraphics[width=0.7\columnwidth]{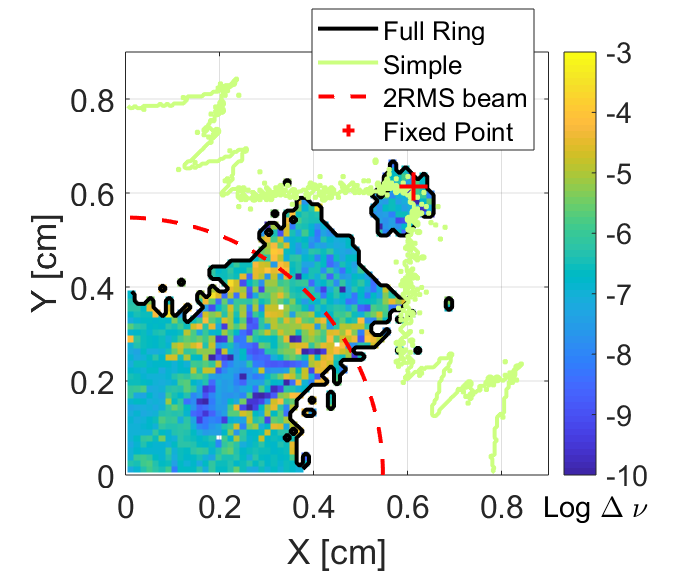}
\label{fig:fma1-xy}
}
\subfigure[Snapshot of beam distribution at beam waist.]{
\includegraphics[width=0.7\columnwidth]{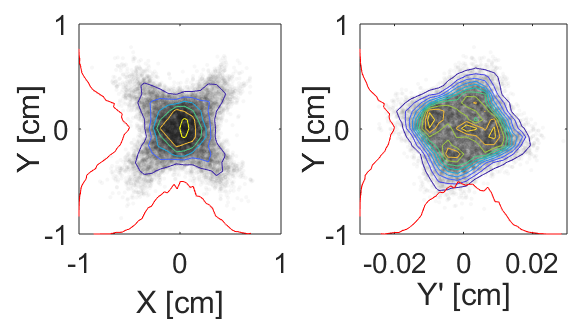}
\label{fig:fma1-dist}
}
\caption{WARP PIC calculation for QIO lattice at peak field $50\ \text{T}/\text{m}^3$ ($\kappa=3984\ m^{-1}$) and fractional tune 0.26 after1024 turns. In \ref{fig:fma1-xy} and \ref{fig:fma1-tune}, results are compared to the equivalent simple model results.}
\label{fig:fma1}
\end{figure}

Results are shown in Fig. \ref{fig:fma1}. 
There is very poor agreement with the predictions of an equivalent simple model, including severe loss of previously stable areas that reduces the spread of tune.
While the lattice tune is far from the condition for quasi-integrability, the greater effect is the nearness of the lattice tune to the fourth order $\nu_x,\nu_y=0.25$ resonances, which are strongly driven by the octupole term. 
Figure \ref{fig:fma1-dist} depicts clear fourth-order resonant structure in the beam distribution. 
Loss of orbit stability appears to coincide with the overlap of the fourth order resonance with nearby sum resonances $3\nu_x+1\nu_y=1$, $1\nu_x+3\nu_y=1$  and  $2\nu_x+2\nu_y=1$.

Studies at other operating points showed similar losses near $\nu_x=0.25$. In one case with fractional tune $\nu_x = 0.24$, the resonant losses resulted in hollowing of the beam core without disrupting stable orbits at higher amplitudes. 
Limiting the tune below $\nu=0.25$ avoids these losses, at the detriment of reducing the induced tune spread. Due to the high-emittance test beam and fixed aperture, it is not possible to operate sufficiently far above $\nu=0.25$. Effectively, the fourth order resonance acts as an upper bound for operating points for single-insert experiments at UMER.

\subsection{Tolerances for closed-orbit distortion and particle phase errors}

A primary concern for the QIO UMER experiments is the sensitivity of the nonlinear lattice relative to UMER tolerances as they are known. 
The two biggest areas of concern are transverse misalignment and errors in the linear focusing function. The simple model described above (octupole insert and thin-lens focusing kick) was used to interrogate the effect of beam misalignment and incorrect linear optics. This approach is based on the assumption that the optics are entirely linear outside of the insert region, and therefore the error-driven dynamics will not accrue during periods of linear transport. High order effects in the linear focusing section (ie, nonlinear space charge or nonlinear field components) are omitted here.
This section summarizes the cases studied; a more complete description of the approach is found in Reference \cite{KierstenThesis}.

\begin{figure}
\centering
\subfigure[Radial aperture vs. orbit error. Dashed line is 80\% aperture.]{
\includegraphics[width=0.225\textwidth]{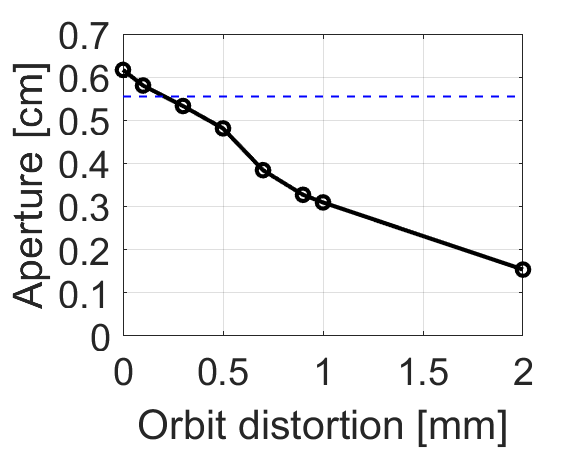}
\label{fig:da-orbit}
}
\subfigure[Radial aperture vs. background field strength. Dashed line is 80\% aperture.]{
\includegraphics[width=0.225\textwidth]{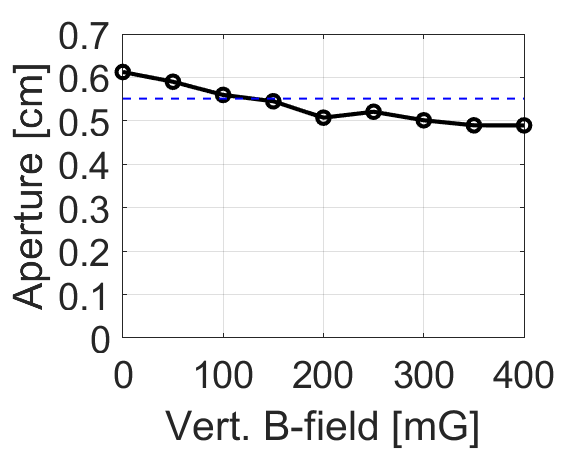}
\label{fig:da-bgfield}
}
\hspace{.5in}
\subfigure[Radial aperture and invariant conservation vs. tune error. Invariant is calculated for a subset of low-amplitude stable particles.]{
\includegraphics[width=0.47\textwidth]{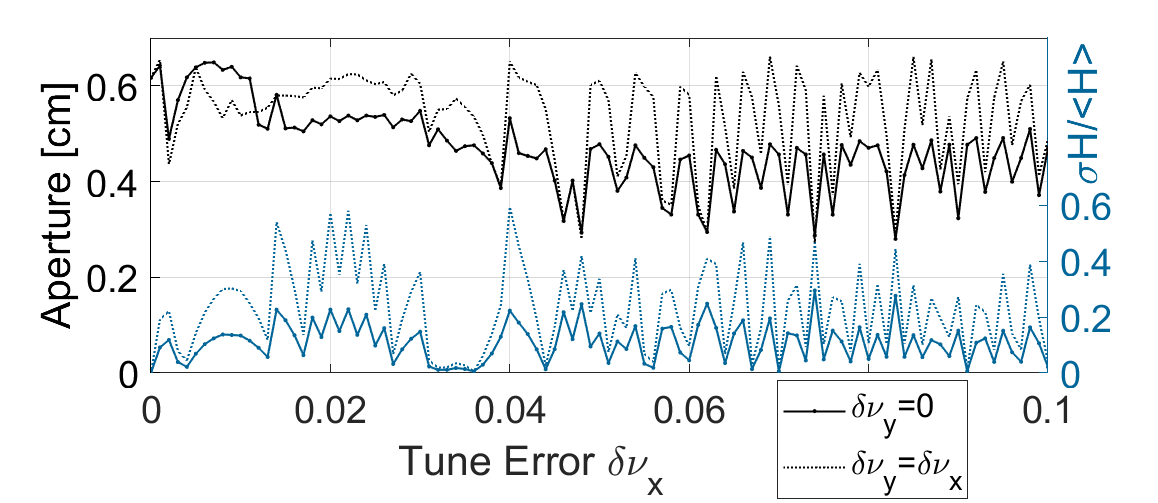}
\label{fig:da-tune}
}
\caption{Effect of steering/focusing errors on QIO lattice stability in simple model with $\beta_*=0.3$ m, peak octupole strength $50\ T/m^3$. Aperture limit is defined by largest circular area that contains only stable particles.}
\label{fig:da}
\end{figure}

First, gross misalignments of the channel were considered. This is the case in which the channel has a tilt with respect to the closed orbit. The amplitude of distortion is parameterized as the transverse offset $\Delta x$ of the closed orbit at insert entrance. The total dynamic aperture decreases dramatically with increasing closed orbit distortion, shown in \ref{fig:da-orbit}, as particles are placed beyond the boundary of stability. Orbit distortions $<0.2$ mm are required to preserve $80\%$ dynamic aperture by area. 
The best orbit control measured over a single $20^\circ$ UMER section is reported in Table \ref{tab:RC9orbit}. Clearly, the measured distortion far exceeds the desired tolerance, particularly in the horizontal plane.

\begin{table}[tb]
\caption{Measured centroid position (in millimeters) at proposed octupole insert location on first turn, measured from center of $20^\circ$ section using first-turn quadrupole response data.} 
\label{tab:RC9orbit}
\centering
\begin{ruledtabular}
\begin{tabular}{lcccc}
Axis & -24 cm & -8 cm & 8 cm & 24 cm \Tstrut\Bstrut \\
\hline 
X  & $-0.18\pm0.16$ & $0.69\pm1.11$ & $-0.14\pm0.49$ & $-0.56\pm0.41$ \Tstrut\\
Y  & $ 0.10\pm0.03$ & $0.05\pm0.99$ & $-0.08\pm0.54$ & $-0.06\pm0.88$ \\
\end{tabular}
\end{ruledtabular}
\end{table}

In general the observed orbit distortion exceeds the desired threshold, especially in the horizontal plane. The main culprit are not mechanical errors, which were corrected and surveyed prior to these measurements, but rather an effective misalignment due to ambient magnetic fields, which provides $\sim 20 \%$ of the total horizontal bending in the ring.
As shown in Fig. \ref{fig:da-bgfield}, adding a constant bending field to the simple orbit-error model shows a weak dependence up to 400 mG, which is the average vertical background field in the ring. For $80\%$ of the dynamic aperture to be preserved, background fields $<100 mG$ (in one plane) are desired.

The simple model was also used to interrogate dependence on focusing errors in the linear-focusing lattice. Assuming single-particle dynamics dominate in the linear sections, a phase shift $\sigma_x = n\pi + \delta \sigma_x$ was added to the thin kick transfer matrix between insertions:
\begin{equation} \normalsize
T_x = 
\begin{bmatrix} 
\cos{\sigma_x} - \alpha_x \sin{\sigma_x } & \beta_x \sin{\sigma_x } \\
-\frac{1-\alpha_x^2}{\beta_x}\sin{\sigma_x }-\frac{2\alpha_x}{\beta_x}\cos{\sigma_x } & \cos{\sigma_x}-\alpha_x\sin{\sigma_x }\\
\end{bmatrix}
\label{eq:tune-error-matrix}
\end{equation} \normalsize
\noindent where $\delta \sigma_x = 0$ for the ideal case, and $\alpha_x$, $\beta_x$ are the Courant-Snyder parameters. The phase error in the $y$-plane is applied in the same way. 

The effect of tune errors on dynamic aperture has a weaker dependence than the centroid errors, as shown in Figure \ref{fig:da-tune}. In general, lattice tunes within $<0.02$ of the optimal value preserved $80\%$ of the stable aperture (by radius), and errors out to $0.1$ resulted in, on average, $30\%$ loss but without a strong dependence on error magnitude. 
On the other hand, conservation of the invariant quantity depends more strongly, with variations $>20\%$ for single-plane tune errors $>0.01$.  In general, it was found that a ``symmetric error," in which $\nu_x -\nu_y$ is small, had less effect than errors with aymmetric tunes. This is consistent with results discussed in Reference \cite{Webb2014}, in which equal chromatic errors are noted to preserve the required symmetry in the invariant quantities.

Finally, replacing the ``ideal" octupole insert (pure octupole field with perfect $\beta^{-3}(s)$ dependence) with the field ``as designed" in Section \ref{sec:octu} (generated by Biot-Savart integration) showed very marginal decrease in stable aperture  ($<1\%$) and increase in the averaged fluctuation of $H_N$ ($+1\%$).

In brief, analysis of a simple octupole lattice shows strong sensitivity to orbit distortion that exceed the current threshold of demonstrated orbit control in UMER. This has motivating the installation of additional steering compensation coils to cancel the ambient fields. 
The sensitivity to lattice tune errors is less crucial (to the levels investigated). Although operating point should be tuned as close to the QI condition as possible, the relatively relaxed tolerance is promising for extension to space-charge dominated beams with intrinsic tune spread.

\section{Discussion and Conclusion}  \label{sec:discuss}


This paper presents an overview of the progress and prospects for testing QIO at UMER. Section \ref{sec:design} described a configuration that supports invariant-conserving orbits at tune spreads up to $\Delta \nu =0.13$ (RMS $0.015$) with a single octupole insert.
The primary limitation of the design as presented is the strength of the predicted effect, which is much lower than the fully-integrable system (up to unity tune spread with multiple insertions \cite{Antipov2017}). This is partially due to the high-density, many-bend lattice that required a perhaps shorter-than-optimal nonlinear insert. For applications beyond proof-of-principle, more insertions are desired.

Some of the technical challenges confronted during this design process are specific to UMER. These include the large orbit distortion and unshielded ambient magnetic field at the time of this work. Simulation-based predictions of sensitivity has motivated the installation of ambient-field-cancelling corrector coils (that are near completion at the time of this submission).
Finally, although many quadrupoles are available for optimizing linear focusing, the lack of dispersion matching at injection limits control of dispersion in the octupole section. In general low energy spread is expected for both coasting-beam and RF-confined experments ($\delta p \sim 0.001$), but the momentum acceptance should be examined in both simulation and experiment.

A consideration for all QIO machines is the apparent sensitivity to the fourth-order resonance stop-band, observed in full-ring PIC simulations although weak in the simple (and highly nonlinear) model.
This limits single-insert experiments to fractional tunes $<0.2$, which correspondingly limits the maximum induced tune spread. 
While it is unsurprising that the octupole resonance is strongly driven, further analysis of the resonance lanscape and sensitivity in a given lattice is necessary to balance decoherence time with amplitude of the driving term for other low-order resonances. Finally, the need to operate with particles \textit{near} the dynamic aperture limit, which corresponds with highly tune-shifted orbits, is in contrast to the typical approach, in which all optics are optimized to increase dynamic aperture. This motivates the need for tightly controlled closed orbit in the insertion region.

Encouragingly, some design aspects were much easier to reconcile than initially expected. 
For a high emittance beam we require modest octupole strength well below the upper limits (determined by threshold for heat-damage to the PCB circuits). 
Also, the purity and precision of the octupole channel does not appear to have a large effect on the dynamics; a model with Biot-Savart generated gridded fields showed only mild degradation of invariant conservation and stable aperture. 
Finally, the system has weak dependence on errors in lattice phase advance, which is promising with respect to internal tune spread due to space charge.
Continuing to advance understanding of the dependence of invariants and stability on space charge will be important in determining the future of integrable optics as a method to improve transport of high intensity beams.

The design presented here will be the starting point for initial experimental runs.
Significant flexibility has been preserved in the linear-focusing optics, which will accomodate operating points with higher fractional tunes as well as multi-insert experiments.
Observing stable beam transport at the design operating point, in tandem with extending the simulation studies to include additional errors and driving terms, will continue to build intuition for design and operation of  integrable optics machines. 
Ultimately, the goal of the UMER QIO program is to extend into denser space charge regimes (ie, row 2 in Table \ref{tab:umer-beams}), following demonstration at the $\mu$A level. 
Although this paper only considered transport in the lowest-charge case, the apparatus as designed supports examination of the transition from the single-particle regime described by the NLIO theory to that in which space charge becomes a significant factor.

\subsection{Acknowledgments}
Thanks to Santiago Bernal for developing low-current beam capability at UMER, as well as a very careful proof-reading of this manuscript. Also many thanks to Levon Dovlatyan, Rami Kishek, Dave Sutter, Eric Montgomery for helpful discussions and technical expertise.
This project has benefited greatly from communication with the IOTA program at FNAL.

Funding for the work was provided through DOE-HEP (Award DESC0010301), NSF (Award PHY1414681) and the NSF GRFP program. 
This manuscript has been authored by UT-Battelle, LLC, under Contract No. DE-AC0500OR22725 with the U.S. Department of Energy.

\bibliography{qio-for-umer}

\end{document}